# Field-free spin-orbit torque switching of an antiferromagnet with perpendicular Néel vector

Zhengde Xu[1*], Jie Ren[1*], Zhengping Yuan[1], Yue Xin[1], Xue Zhang[1], Shuyuan Shi[3], Yumeng Yang[1,2] and Zhifeng Zhu[1,2†]

[1]School of Information Science and Technology, ShanghaiTech University, Shanghai, China 201210

[2]Shanghai Engineering Research Center of Energy Efficient and Custom AI IC, Shanghai, China 201210

[3]Fert Beijing Institute, MIIT Key Laboratory of Spintronics, School of Integrated Circuit Science and Engineering, Beihang University, Beijing, China

**ABSTRACT**

The field-free spin-orbit torque induced 180° reorientation of perpendicular magnetization is beneficial for the high performance magnetic memory. The antiferromagnetic material (AFM) can provide higher operation speed than the ferromagnetic counterpart. In this paper, we propose a trilayer AFM/Insulator/Heavy Metal structure as the AFM memory device. We show that the field-free switching of the AFM with perpendicular Néel vector can be achieved by using two orthogonal currents, which provide the uniform damping-like torque and stagger field-like torque, respectively. The reversible switching can be obtained by reversing either current. A current density of $1.79 \times 10^{11}$ A/m$^2$ is sufficient to induce the switching. In addition, the two magnetic moments become noncollinear during the switching. This enables an ultrafast switching within 40 picoseconds. The device and switching mechanism proposed in this work offer a promising approach to deterministically switch the AFM with perpendicular Néel vector. It can also stimulate the development of ultrafast AFM-based MRAM.

**INTRODUCTION**

The electrical control of magnetic states provides promising solutions for the non-volatile data storage. Previous research has been focused on the study of ferromagnets (FMs), where the nanosecond switching and the sub-10 nm device have been demonstrated.[1] To further improve the performance of magnetic devices, recent research focus has been shifted from FMs to antiferromagnets (AFMs). Due to the strong exchange interactions in the AFM, it is able to produce ultrafast spin dynamics in the picosecond timescale.[2-5] In addition, since the magnetic moments in the AFM are antiparallelly aligned, this material does not produce stray field, which enables the further scale down of spintronic devices and makes it robust against external magnetic perturbations.[6]

To change the magnetic state of FMs, the current-induced spin-transfer torque (STT) is widely used.[7,8] This stimulates the development of STT-MRAM. However, since the current flows directly through the memory cell, the device is prone to breakdown. To avoid this, it is found that the spin current can also be obtained through the spin-orbit torque (SOT),[9-12] which is even more efficient than the STT. Previous studies have demonstrated that SOT can be classified into the damping-like torque (DLT), $\tau_{DLT}=-\mathbf{m}\times(\mathbf{m}\times\boldsymbol{\sigma})$ and the field-like torque (FLT), $\tau_{FLT}=\mathbf{m}\times\boldsymbol{\sigma}$.[13-16] Here $\boldsymbol{\sigma}$ is the spin polarization. The SOT has been demonstrated to switch both the in-plane[17] and perpendicular magnetization[9,13,18-21] of FMs, where it is commonly known that the DLT is responsible for the switching and the FLT only plays a secondary role.[22] Moreover, utilizing two orthogonal currents to achieve field-free switching in FM has also been demonstrated.[23,24]

In contrast, neither the DLT nor the FLT can switch the magnetization in AFM due to cancellation of torque induced by the antiparallelly aligned magnetic moments. When only the DLT is applied, both magnetic moments experience the same torque, i.e., $\tau_{DLT,A}=\tau_{DLT,B}=-\mathbf{m}_i\times(\mathbf{m}_i\times\boldsymbol{\sigma}_i)$ with $\boldsymbol{\sigma}_A=\boldsymbol{\sigma}_B$. The subscript $i$ denotes different sublattices. This is defined as the uniform DLT, under which the magnetization evolves into oscillation in the plane perpendicular to the spin polarization.[25] When only the FLT is applied, its effect is the same as the magnetic field, resulting in the spin flop and the magnetization is reoriented to be perpendicular to the spin polarization.[26] Recently, it has been demonstrated both theoretically[27,28] and experimentally[29-31] that 90° magnetization reorientation can be achieved in the in-plane AFMs with locally broken inversion symmetry, such as the CuMnAs and $Mn_2Au$. In these materials, the adjacent magnetic moments form inversion partners. When an electrical current is applied, it generates opposite $\boldsymbol{\sigma}$ acting on the different sublattices in the form of FLT, i.e., $\tau_{FLT,A}=\mathbf{m}_A\times\boldsymbol{\sigma}_A$ and $\tau_{FLT,B}=\mathbf{m}_B\times\boldsymbol{\sigma}_B$ with $\boldsymbol{\sigma}_A=-\boldsymbol{\sigma}_B$. This is known as the stagger FLT. Furthermore, some theoretical studies have investigated the current induced switching of AFM with perpendicular Néel vector, such as $Mn_3Sn$.[32,33]

Currently, the FMs are mainly used in the MRAM as the storage element. It has been proposed that the use of FMs with perpendicular magnetization can reduce the critical switching current and enable the further scale down of memory cells.[34] However, the deterministic SOT switching of perpendicular magnetization requires an external magnetic field.[18,35] Many studies have been carried out to realize the field-free switching in the FM system.[36-38] In order to further improve the performance of MRAM, it is desired to replace the perpendicular FM with AFM with perpendicular Néel vector. Therefore, it is important to realize the field-free 180° switching of the AFM with perpendicular Néel vector,[39-45] where the opposite states can be distinguished by the second-harmonic magnetoresistance effect combined with a second-order magnetotransport effect caused by an alternating probing current $\mathbf{J}_{ac}$ along the $\mathbf{x}$ axis[44] or the anomalous Hall effect (AHE).[33,40] Recent studies demonstrated that a giant tunneling magnetoresistance (TMR) effect can be produced in antiferromagnetic tunnel junctions (AFMTJs) in response to the 180° rotation of the Néel vector, which enables efficient readout of the compensated AFM states.[46,47]

In this paper, we propose to apply two currents into an AFM/Insulator/Heavy Metal (HM) heterostructure, where the AFM is required to possess locally broken inversion symmetry. The two currents are perpendicular to each other to produce orthogonal spin polarization. The deterministic 180° switching of the AFM with perpendicular Néel vector can then be achieved by the combined effect of the uniform DLT induced by the HM and the stagger FLT induced by the AFM. Furthermore, the proposed switching mechanism is robust against the current delay. This is similar to the toggle MRAM which is proposed to eliminate the half-selection problem.

More generally, we find that the successful switching can be achieved using two torques with the following properties. Initially, their combined strength should be sufficient to induce the reorientation of magnetic moment. After the magnetic moment is switched to the opposite direction, one of the torques should reverse its direction to balance the other torque, so that the magnetic moment can stay switched. By analyzing the torques experienced by the AFM sublattices in our device, we find that the abovementioned two conditions are satisfied due to the different symmetry of uniform DLT and stagger FLT.

**METHODOLOGY**

The device studied in this paper is shown in Fig. 1(a). It consists of an AFM layer and a HM layer sandwiched by an insulating layer.[48,49] The AFM material consists of two sublattices with local inversion asymmetry. The magnetization in the AFM is assumed to be perpendicular to the film plane. Two currents are applied to the device. $\mathbf{J}_{c1}$ is injected into the HM layer along the **x** direction, which induces a vertical spin current $\mathbf{J}_s = \theta_{SH}\boldsymbol{\sigma}\times\mathbf{J}_c$ due to the spin-Hall effect (SHE). Here $\theta_{SH} = -0.1$ denotes the spin Hall angle. The insulating layer is used to electrically isolate the AFM and HM layers but is transparent to the pure spin current $\mathbf{J}_s$, which can pass through the insulating layer and then acts on the two sublattices in the form of the magnon transfer torque,[48-50] i.e., a uniform DLT $[-\mathbf{m}_i\times(\mathbf{m}_i\times\boldsymbol{\sigma}_i)]$ with $\boldsymbol{\sigma}_A=\boldsymbol{\sigma}_B$. Note that the direction of DLT is the same for the two sublattices. This is easily understood since the DLT is even in **m**. For example, when $\mathbf{J}_{c1}$ is along the −**x** direction, the SHE gives $\boldsymbol{\sigma} = -\mathbf{y}$. Therefore, both $\mathbf{m}_A$ and $\mathbf{m}_B$ experience a DLT pointing in the +**y** direction. When $\mathbf{J}_{c1}$ is further increased, $\mathbf{m}_A$ and $\mathbf{m}_B$ are not able to maintain the original state, and it oscillates in the plane perpendicular to the spin polarization, i.e., the **xz** plane [see Fig. 1(b)].[25,51]

On the other hand, when $J_{c1}$ is removed and an orthogonal current $J_{c2}$ is applied to the AFM layer along the **y** direction, due to the local broken symmetry, the sublattices experience stagger FLT ($\mathbf{m}_i \times \boldsymbol{\sigma}_i$) in which $\boldsymbol{\sigma}$ is opposite, i.e., $\boldsymbol{\sigma}_A = +\mathbf{z} \times \mathbf{J}_{c2}$ and $\boldsymbol{\sigma}_B = -\mathbf{z} \times \mathbf{J}_{c2}$.[27] Under a large $J_{c2}$, both $\mathbf{m}_A$ and $\mathbf{m}_B$ are stabilized in the direction along the spin polarization (i.e., the **x** direction).[27,28,30,31] The corresponding magnetization dynamics is illustrated in Fig. 1(c).

We adopt a macrospin model to describe the magnetic dynamics. For simplicity, it neglects the atomic-scale variation, such as the domain wall or nucleation switching process. We also consider the FLT generated by the HM layer and the DLT in the AFM layer and we setup simulation with the strength of $\tau_{FLT}/\tau_{DLT} = 0.3$[52,53] in the HM layer and $\tau_{DLT}/\tau_{FLT} = 0.05$[28] in the AFM layer. The addition of these torques does not affect the deterministic switching of AFM with perpendicular Néel vector proposed in this work.

In the dynamic simulation, we first set the initial direction of magnetic moments $\mathbf{m}_A$ and $\mathbf{m}_B$ along +**z** or −**z**, respectively. The dynamics of the sublattice moments **m** is described by the coupled Landau-Lifshitz-Gilbert (LLG) equations,

$$\frac{d\mathbf{m}_i}{dt} = -\gamma_i \mathbf{m}_i \times \mathbf{H}_{\text{eff},i} + \alpha_i \mathbf{m}_i \times \frac{d\mathbf{m}_i}{dt} - \gamma_i \tau_{\text{DLT},i} \mathbf{m}_i \times (\mathbf{m}_i \times \boldsymbol{\sigma}_i) + \gamma_i \tau_{\text{FLT},i} \mathbf{m}_i \times \boldsymbol{\sigma}_i$$

The four terms on the right hand side represent the precession torque ($\tau_{\text{pre}}$), the Gilbert damping torque ($\tau_{\text{damping}}$), the DLT ($\tau_{\text{DLT}}$) and the FLT ($\tau_{\text{FLT}}$), respectively. $\gamma_i$ is the gyromagnetic ratio and $\alpha_i$ is the damping constant. In this study, we set $\gamma_A = \gamma_B$ and $\alpha_A = \alpha_B = 0.005$.[51] The effective field, $\mathbf{H}_{\text{eff}}$, consists of the crystalline anisotropy with the anisotropy constant $k_u = 0.196$ meV[32] and the exchange constant with $A_{\text{ex}} = 3.74 \times 10^{-21}$ J.[32] We also consider the thermal fluctuations and Oster field in the effective field calculation. However, there is a negligible effect on the results which is attributed to the robustness of AFMs. The strength of DLT is described by

$$\tau_{\text{DLT},i} = \frac{\hbar}{2} \frac{J_{s,1}}{et_{\text{HM}} M_{s,i}} \frac{\gamma_i}{(1+\alpha_i^2)},$$

where $\hbar$ is the reduced Planck constant, $e$ is the electron charge and $M_{s,A} = M_{s,B} = 1.2 \times 10^4$ A/m is the saturation magnetization.[54] The strength of FLT is similarly given by

$$\tau_{\text{FLT},i} = \frac{\hbar}{2} \frac{J_{s,2}}{et_{\text{AFM}} M_{s,i}} \frac{\gamma_i}{(1+\alpha_i^2)}.$$

[55-57] The Runge-Kutta fourth-order method[58] has been used to solve the LLG equation.

**RESULTS AND DISCUSSION**

Compared to the magnetic materials with in-plane magnetization, the use of materials with perpendicular magnetization ensures a lower switching current, higher thermal stability and smaller footprint.[34,59] Therefore, it is desirable to achieve deterministic switching of the perpendicular Néel order parameter, **l**=(**m**$_A$−**m**$_B$)/2.

After reproducing the well known results by applying **J**$_{c1}$ and **J**$_{c2}$ separately in Fig. 1(b) and Fig. 1(c), we further show that the deterministic switching of the perpendicular Néel order parameter can be achieved by simultaneously applying **J**$_{c1}$ and **J**$_{c2}$. As shown in Fig. 2(a), when **J**$_{c1}$ is applied in the −**x** direction and **J**$_{c2}$ is applied in the +**y** direction, **l** is switched from up to down. The switching trajectory is illustrated in Fig. 2(b). The switching starts with a fast reorientation of **m**, followed by the precession of **m** in a small angle, and finally **m** is stabilized in the opposite direction. This switching trajectory is very similar to that in the SOT induced switching of the perpendicular FM.[60] However, the switching of perpendicular FM requires an external field to break the symmetry. In contrast, we use the combined effect of two orthogonal currents to realize field-free switching. In addition, we notice that the switching is completed within 40 ps, which is two orders faster than that in the ferromagnetic devices. We also find that **m**$_A$ and **m**$_B$ are not always collinear during the switching. As shown in the right **y** axis of Fig. 2(a), the maximum angle difference is 9.14°. This noncollinearity gives rise to a strong exchange field, leading to the ultrafast switching.[3] Similarly, the down to up switching can be achieved by reversing either **J**$_{c1}$ or **J**$_{c2}$. Fig. 2(c) and 2(d) show the magnetization dynamics when **J**$_{c1}$ is applied in the +**x** direction and **J**$_{c2}$ is remained in the +**y** direction.

The observed AFM switching can be understood by analyzing the torques experienced by the sublattices. For example, as shown in Fig. 3(a), when **m**$_A$ is initialized in the +**z** direction, the **J**$_{c1}$ in the −**x** direction induces $\tau_{DLT,A}$ pointing in the −**y** direction [i.e., −**m**$_i$×(**m**$_i$×**σ**$_i$) =−**z**×(**z**×−**y**)=−**y**], and the **J**$_{c2}$ in the +**y** direction induces $\tau_{FLT,A}$ also in the −**y** direction (i.e., **m**$_i$×**σ**$_i$ =**z**×−**x**=−**y**). As a result, **m**$_A$ is slightly tilted toward the −**y** direction under $\tau_{DLT,A}$ and $\tau_{FLT,A}$. Apply the same analysis to **m**$_B$ and one can find that **m**$_B$ is also tilted toward the −**y** direction. This gives rise to an exchange field acting on **m**$_A$ (**H**$_{ex,A}$=−λ**m**$_B$) pointing in the +**y** direction. Since **H**$_{ex,A}$ is much larger than the anisotropy field **H**$_{an,A}$ in the **H**$_{eff}$, we then consider the effect of precession torque [$\tau_{pre,A}$=−**m**$_A$×**H**$_{eff,A}$=−**z**×(+**y**)=+**x**],[61] which is orthogonal to **m**$_A$ and drags **m**$_A$ from +**z** toward +**x** inside the **xz** plane. After that, **m**$_A$ continues to rotate and finally stabilized in the −**z** direction. This completes the 180° switching of **m**$_A$ from up to down. After **m**$_A$ is switches to the −**z** direction, as shown in Figure 3(b), the direction of $\tau_{FLT,A}$ is reversed, whereas $\tau_{DLT,A}$ remains in the original direction. They are in the opposite direction and compensating each other. Thus, under the negative $J_{c1}$ and positive $J_{c2}$, $m_A$ is switched from +z to −z, in which it has the minimum energy and the system rests in the equilibrium state. The same analysis can be applied to $m_B$, which is switched from down to up under the same stimulus. As a result, the Néel order parameter is switched from up to down.

Similarly, to switch $m_A$ from −z to +z, one can reverse the $J_{c1}$ from −x to +x direction and keep the $J_{c2}$ in the +y direction [see Fig.3(c)]. Under this condition, $J_{c1}$ gives rise to a $\tau_{DLT,A}$ pointing in the +y direction, and $J_{c2}$ gives rise to a $\tau_{FLT,A}$ which is also in the +y direction. Therefore, the exchange field acting on $m_A$ is pointing in the −y direction, leading to a precession torque drags $m_A$ from −z to +x and finally stabilized in the +z direction, where $\tau_{DLT,A}$ and $\tau_{FLT,A}$ are balanced [see Fig. 3(d)]. This completes the switching of Néel order parameter from down to up. The different switching conditions have been summarized in Table 1. The two cases discussed before correspond to the third (switching of l from +z to −z) and first row (switching of l from −z to +z), respectively. Note that the trajectory of $m_A$ in both cases passes through +x due to the precession torque. To complete the table, one can fix the direction of $J_{c2}$ in the −y direction. The switching of l from +z to −z can then be achieved by setting the $J_{c1}$ in the +x direction, whereas the opposite switching can be realized by reversing $J_{c1}$. Note that the trajectory of $m_A$ now passes through −x as shown in the last column of Table 1.

From these discussions, one notices that there exists a competition between $\tau_{DLT}$ and $\tau_{FLT}$, which are originated from $J_{c1}$ and $J_{c2}$, respectively. To understand the complete magnetization dynamics, we then study the switching phase diagrams (SPD). As shown in Fig. 4, different dynamics regions are observed when $J_{c1}$ and $J_{c2}$ are varied. When either $J_{c1}$ or $J_{c2}$ is small, the magnetization remains in the original state since the $\tau_{DLT}$ or $\tau_{FLT}$ alone is not sufficient to overcome the anisotropy. This is denoted by the yellow region in the middle of SPD. On the other hand, when only one of $\mathbf{J}_{c1}$ and $\mathbf{J}_{c2}$ is large, AFM switching cannot be achieved. This has been verified in Fig. 1(b) and 1(c), i.e., when only $\mathbf{J}_{c1}$ is large, the Néel order parameter evolves into the oscillation in the plane perpendicular to the spin polarization; when only $\mathbf{J}_{c2}$ is large, the Néel order parameter is aligned with the direction perpendicular to $\mathbf{J}_{c2}$.[27,29] The AFM switching (the blue region) can only occur when both $\mathbf{J}_{c1}$ and $\mathbf{J}_{c2}$ are large, so that $\tau_{DLT}$ and $\tau_{FLT}$ can be added up in the beginning to overcome the anisotropy. After the Néel order parameter is switched to the opposite direction, $\tau_{FLT}$ is reversed since it is odd in $\mathbf{m}$, resulting in a balanced $\tau_{DLT}$ and $\tau_{FLT}$ (Fig. 3). Thus, the Néel order parameter can remain in the switched direction. In addition, the blue region in the SPD can be divided into two cases. When $\mathbf{J}_{c1}$ ranges from $1 \times 10^{10}$ A/m$^2$ to $6.5 \times 10^{10}$ A/m$^2$, the critical $\mathbf{J}_{c2}$ required for switching decreases as $\mathbf{J}_{c1}$ increases. This also supports our explanation that one needs the addition of $\tau_{DLT}$ and $\tau_{FLT}$ in the beginning to initiate the reorientation of Néel order parameter. However, when $\mathbf{J}_{c1}$ is larger than $6.5 \times 10^{10}$ A/m$^2$, the critical $\mathbf{J}_{c2}$ increases with $\mathbf{J}_{c1}$. Under a very large $\mathbf{J}_{c1}$, although the initial torque is sufficient to overcome the anisotropy to start the AFM switching, it also requires a comparable $\mathbf{J}_{c2}$ so that the Néel order parameter can be stabilized in the opposite direction. Otherwise, the Néel order parameter will evolve into oscillation as shown in the edge green region of the SPD. Therefore, the SPD shown in Fig. 4 clearly illustrate our proposed picture of 180° deterministic switching of AFM with perpendicular Néel vector, i.e., the two torques should add up to initialize the magnetization switching. After $\mathbf{l}$ is switched to the opposite direction, one of the torques is required to be reversed to balance the other torque, so that

**l** can stay switched.

**CONCLUSION**

In summary, we study the spin-orbit torque-induced 180° magnetization switching in the AFM with perpendicular Néel vector/Insulator/Heavy Metal heterojunction. When only the uniform DLT or the stagger FLT is applied, the Néel order parameter develops into oscillation or reorients to the perpendicular direction, respectively. In contrast, when the uniform DLT and the stagger FLT are applied simultaneously, field-free 180° magnetization switching of AFM with perpendicular Néel vector can be achieved. By analyzing the torques experienced by the sublattices and the switching phase diagram, we conclude that the switching is initiated by the addition of the two torques in the beginning to overcome the anisotropy, and then one of the torques reverses to balance the other torque after the Néel order parameter is switched to the opposite direction. The abovementioned requirements are satisfied by the different symmetry of uniform DLT and stagger FLT. Furthermore, the switching completes in the picosecond range using moderate current density. Our study of field-free switching as well as the switching of perpendicular magnetization are beneficial for the development of next generation AFM-based ultrafast MRAM.

†Corresponding Author: zhuzhf@shanghaitech.edu.cn

*These authors contributed equally to this work.

**ACKNOWLEDGMENTS**: Authors Zhifeng Zhu and Yumeng Yang received funding from National Key R&D Program of China (Grant No. 2022YFB4401700). Author Zhifeng Zhu received funding from Shanghai Sailing Program (Grant No. 20YF1430400), and National Natural Science Foundation of China (Grants No. 12104301). Author Yumeng Yang received funding from National Natural Science Foundation of China (Grants No. 62074099).

**Figures**

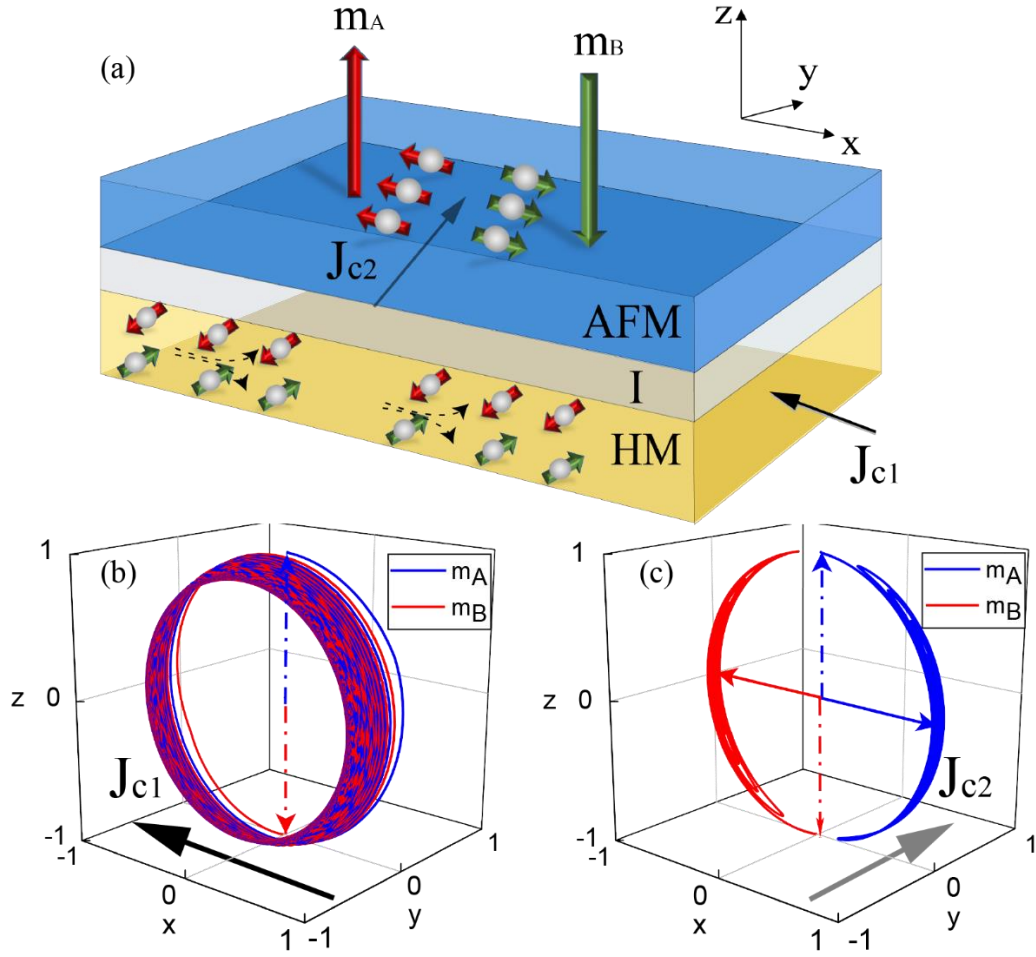

FIG. 1. (a) Device geometry consists of AFM/Insulator/Heavy Metal layers. The dashed arrows indicate the flow of electron spin. The size of AFM layer is 50 nm × 50 nm × 10 nm. The size of insulating layer (NiO) is 50 nm × 50 nm × 3 nm. The size of HM layer is 50 nm × 50 nm × 5 nm. (b) When $\mathbf{J}_{c1} = 1.26\times10^{11}$ A/m$^2$, the magnetization oscillates in the plane perpendicular to the spin polarization, i.e., **xz** plane. (c) When $\mathbf{J}_{c2} = 5.5\times10^{11}$ A/m$^2$, the magnetization switches from the **z** to the **x** direction. The dotted and solid arrow represent the initial and final position of **m**, respectively.

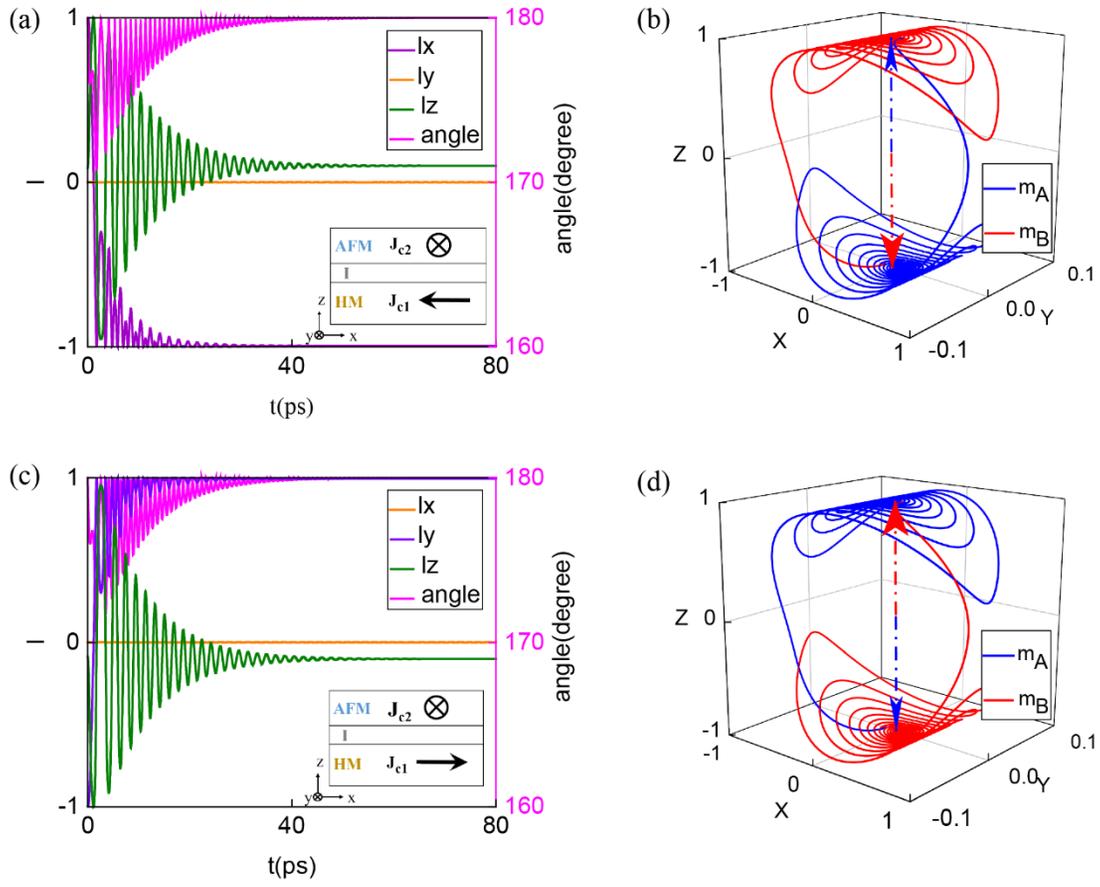

FIG. 2. The time evolution (a) and the 3D switching trajectory (b) of **l** when $\mathbf{J}_{c1} = 7.4 \times 10^{10}$ A/m$^2$ and $\mathbf{J}_{c2} = 1.79 \times 10^{11}$ A/m$^2$ are applied in the −**x** and +**y** directions, respectively. The initial direction of **m** is represented by the dotted arrow. The time evolution (c) and the 3D switching trajectory (d) of **l** when $\mathbf{J}_{c1}$ is changed to the +**x** direction.

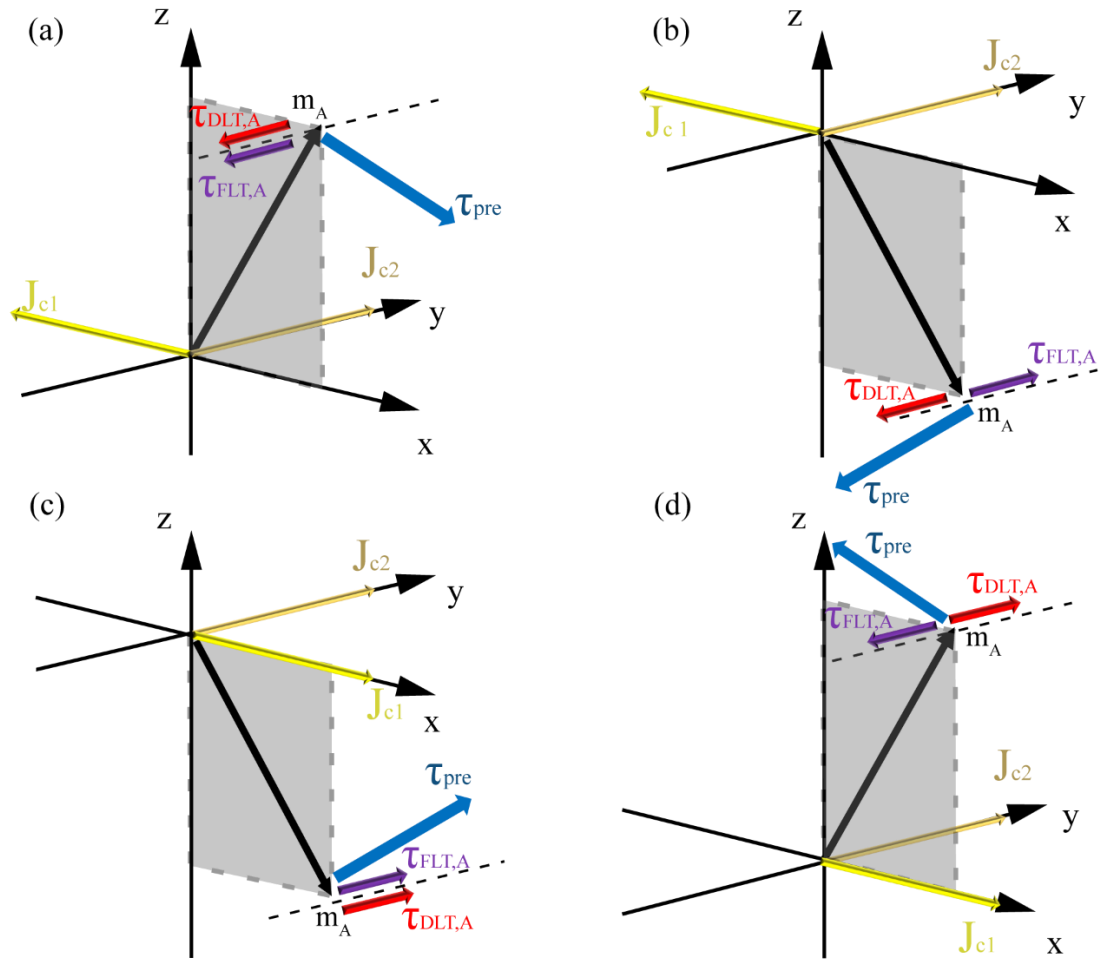

FIG. 3. Torques experienced by $\mathbf{m}_A$ when it is switched from (a) top to (b) bottom, and from (c) bottom to (d) top. $\boldsymbol{\tau}_{pre}$ and $\mathbf{m}_A$ are always orthogonal. The shaded rectangle is used to illustrate the **xz** plane. The dash line is parallel to the **y** axis.

Table 1. The favored magnetization directions under different $\mathbf{J}_{c1}$ and $\mathbf{J}_{c2}$.

| $\mathbf{J}_{c1}$ | $\mathbf{J}_{c2}$ | Favored **l** | Trajectory of $\mathbf{m}_A$ |
|---|---|---|---|
| + | + | +**z** | −**z**→+**x**→+**z** |
| + | − | −**z** | +**z**→−**x**→−**z** |

| - | + | −z | +z→+x→−z |
| - | - | +z | −z→−x→+z |

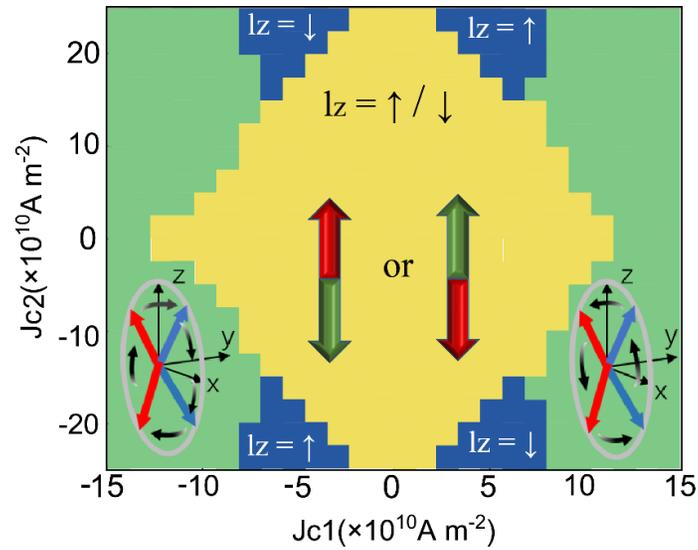

FIG. 4. Switching phase diagram as a function of $\mathbf{J}_{c,1}$ and $\mathbf{J}_{c,2}$. In the yellow region, the magnetization remains in the original state. The blue region denotes the successful 180° switching of the Néel order parameter. In the green region, the magnetization develops into oscillation, where the black arrows illustrate the oscillation direction.